\documentclass[3p,times,twocolumn]{elsarticle}

\usepackage{ecrc}

\volume{00}

\firstpage{1}

\journalname{Nuclear Physics B Proceedings Supplement}

\runauth{}

\jid{nuphbp}

\jnltitlelogo{Nuclear Physics B Proceedings Supplement}

\usepackage{amssymb}

\usepackage[figuresright]{rotating}

\begin{document}

\begin{frontmatter}



\dochead{}

\title{JUNO: A Next Generation Reactor Antineutrino Experiment}


\author{Liang Zhan}

\address{Institute of High Energy Physics, Chinese Academy of Sciences, Beijing 100049, China
\\ \textrm{zhanl@ihep.ac.cn}}

\begin{abstract}
The mass hierarchy and the CP phase are the main focus of the next generation neutrino oscillation experiments. Jiangmen Underground Neutrino Observatory (JUNO), as a medium baseline reactor antineutrino experiment, can determine the neutrino mass hierarchy independent of the CP phase. The physics potential on the mass hierarchy, and other measurements are reviewed. The preliminary design options for a 20~kton detector with an energy resolution of $3\%/\sqrt{E_{vis}}$ are illustrated. The main technical challenges on the PMT and scintillator are discussed and the corresponding R\&D efforts are presented.

\end{abstract}


\end{frontmatter}


\section{Introduction}
\label{introduction}
A large neutrino mixing angle $\theta_{13}$ was observed by the latest reactor~\cite{An:2012eh, Abe:2011fz, Ahn:2012nd, An:2013uza, An:2013zwz} and accelerator~\cite{Abe:2011sj, Adamson:2011qu} neutrino oscillation experiments. In the next few years, the precision of $\sin^2{2\theta_{13}}$ can reach to 3\% by continuous running of Daya Bay. The remaining unknown parameters are the neutrino mass hierarchy (the sign of $\Delta m^2_{31}$ or $\Delta m^2_{32}$) and the CP phase, which are the focus of next generation neutrino oscillation experiments. One possibility to determine the mass hierarchy is to measure the mass effect which is different for normal hierarchy and inverted hierarchy, and hence atmospheric~\cite{INO, PINGU} and accelerator~\cite{HK, LBNE, LBNO} neutrino experiments are proposed.  Another possibility is raised~\cite{Petcov:2001sy, Choubey:2003qx, Learned:2006wy, Zhan:2008id, Zhan:2009rs} to use the oscillation interference effect between $\Delta m^2_{31}$ and $\Delta m^2_{32}$ by reactor antineutrinos.

\begin{figure*}
\begin{center}
\begin{tabular}{c}
\includegraphics*[ width=0.8\textwidth]{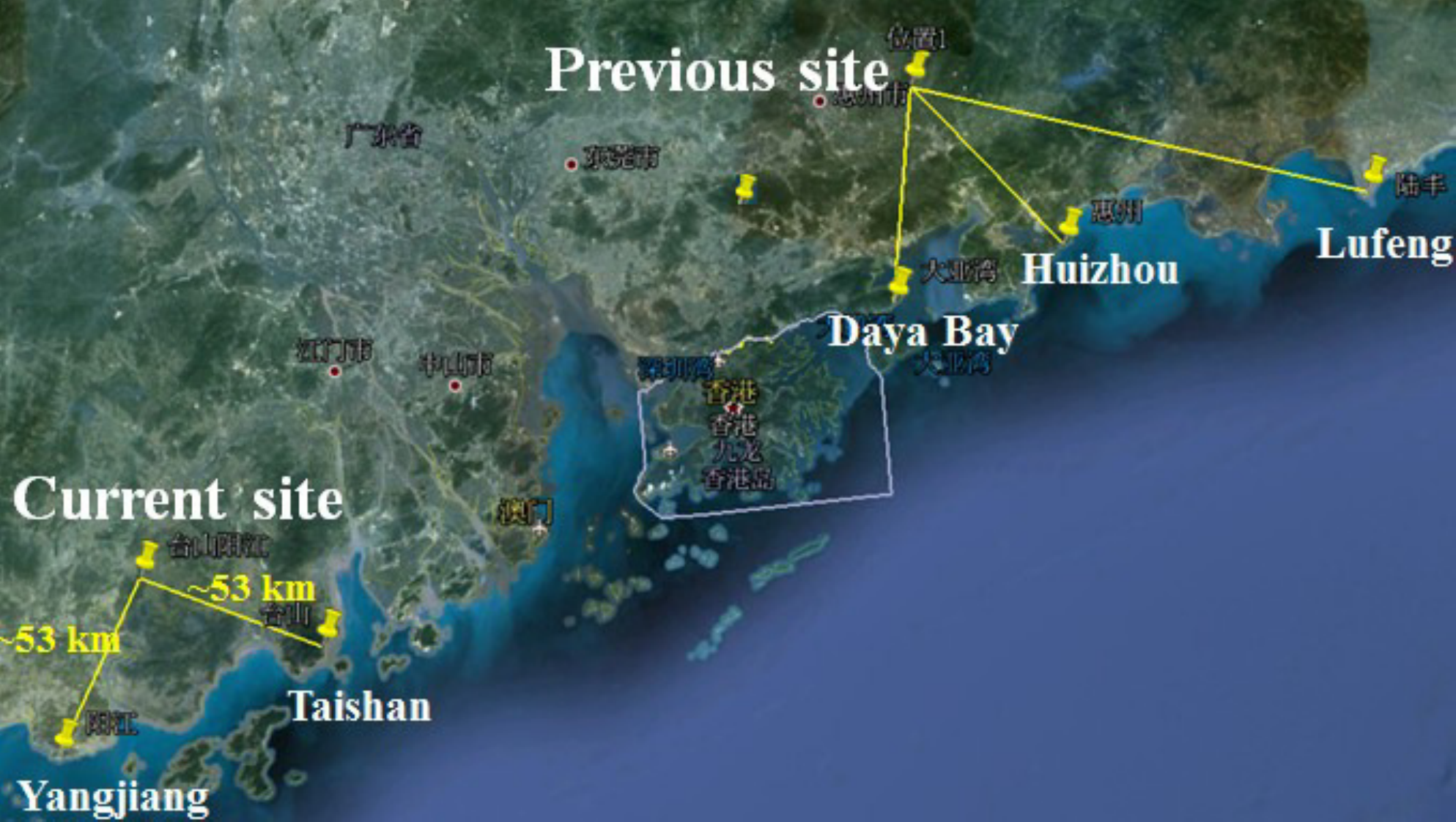}
\end{tabular}
\end{center}
\caption{Candidate site of the JUNO experiment. \label{fig:site}}
\end{figure*}

The Jiangmen Underground Neutrino Observatory (JUNO) uses 20~kton liquid scintillator (LS) as detector target. The primary goal is to determine the mass hierarchy using the reactor antineutrinos by precise measuring the oscillation spectrum with a $3\%/\sqrt{E_{vis}}$ energy resolution. A candidate site is located at Jiangmen in South China, with baselines around 53~km from Taishan and Jiangmen nuclear power plants, shown in Figure~\ref{fig:site}.
In addition to the mass hierarchy, such a large and high precision detector has other rich physics potentials. The  neutrino mixing parameters ($\Delta m^2_{21}$, $\Delta m^2_{31}$ and $\sin^2\theta_{12}$) can be measured with a precision better than 1\%. JUNO detector is also sensitive to supernova neutrinos, geo-neutrinos, solar neutrinos and atmospheric neutrinos, and can probe the sterile neutrinos and exotic physics like nucleon decays.

\section{Sensitivity to mass hierarchy}
At the candidate site, the baselines and the reactor power of the nuclear power plants are listed~\cite{JUNO} in Table~\ref{tab:reactors}.
\begin{table}
\centering
\begin{tabular}{|c|c|c|c|c|c|c|}\hline\hline
Cores & TS-C1 & TS-C2 & TS-C3 & TS-C4 \\
\hline Power (GW) & 4.6 & 4.6 & 4.6 & 4.6  \\ \hline
Baseline(km) & 52.76 & 52.63 & 52.32 & 52.20  \\
\hline\hline
Cores & YJ-C1 & YJ-C2 & YJ-C3 & YJ-C4 \\
\hline Power (GW) & 2.9 & 2.9 & 2.9 & 2.9\\ \hline
Baseline(km) & 52.75 & 52.84 & 52.42 & 52.51  \\
\hline\hline
Cores &  YJ-C5  & YJ-C6  & DYB & HZ\\
\hline Power (GW) &  2.9 & 2.9 & 17.4 & 17.4\\ \hline
Baseline(km) &  52.12 & 52.21 & 215 & 265\\
\hline
\end{tabular}
\caption{List of the baselines and reactor power for the
Yangjiang (YJ) and Taishan (TS) reactor power plants, as well as the remote reactors of Daya Bay (DYB) and
Huizhou (HZ), at the candidate site.}
\label{tab:reactors}
\end{table}
The Yangjiang and Taishan reactor power plants are under construction and will be completed around 2020. The site was selected to minimize the baseline difference between Yangjiang and Tainshan reactor cores for better sensitivity, which will be discussed later. The survival probability of the reactor antineutrinos can be written as
\begin{eqnarray}
\label{eq:Pee}
 P_{ee} = 1 &-&  \cos^4\theta_{13}\sin^22\theta_{12}\sin^2\Delta_{21} \nonumber \\
 &-& \cos^2\theta_{12}\sin^22\theta_{13}\sin^2\Delta_{31}\nonumber\\
 &-& \sin^2\theta_{12}\sin^22\theta_{13}\sin^2\Delta_{32},
\end{eqnarray}
where $\Delta_{ij} = 1.27 \Delta m^{2}_{ij}L/E$, $\Delta m^{2}_{ij}$ is neutrino mass-squared difference ($m^2_i - m^2_j$) in eV$^2$, $\theta_{ij}$ is neutrino mixing angle,  $L$ is the baseline length from reactor to $\overline{\nu}_e$ detector in meter, and $E$ is the $\overline{\nu}_e$ energy in MeV. The reactor antineutrinos can be detected in liquid scintillator detector via coincident signals of inverse $\beta$ decay (IBD), $\bar{\nu_e} + p \rightarrow e^+ + n$. More than hundred thousand IBD signals can be collected by 6 year running, assuming an 80\% effective detection efficiency (considering event selection efficiency, live time ratio and reactor full power ratio).

To calculate the sensitivity of mass hierarchy, we fit the simulated spectrum to the normal hierarchy (NH) and inverted hierarchy (IH) cases respectively using a $\chi^2$ function,
\begin{eqnarray}
\label{eq:chi2}
\chi^2 =\sum_i  \frac{\left(T_i - F_i(1 + \epsilon + \epsilon_i)\right)^2}{T_i}  \nonumber  \\
+ \sum_p  \frac{(p_f-p_t)^2}{\sigma_p^2} + \sum_i  \frac{\epsilon_i^2}{\sigma_i^2},
\end{eqnarray}
where $T_i$ is the input antineutrino spectrum assuming the truth is NH or IH, $F_i$ is the fitted spectrum. The $\epsilon$ is the normalization factor of the antineutrino spectrum and it is floating in the $\chi^2$ function. There is no constrain on $\epsilon$ and it means we only use the shape information.  The $\epsilon_i$ is uncorrelated between bins and so the shape can fluctuate. The $p$ in $\chi^2$ function denotes the oscillation parameters, including $\sin^22\theta_{13}$, $\sin^2\theta_{12}$, $\Delta m^2_{21}$, and $\Delta m^2_{ee}$ (an effective mass-squared difference defined as a combination of $\Delta m^2_{31}$ and $\Delta m^2_{32}$~\cite{JUNO}). The $p_f$ and $p_t$ are fitted and input oscillation parameters respectively. The $\sigma_p^2$ is the uncertainty of the corresponding oscillation parameter. As expected, the spectrum can fit the input spectrum better if using the truth mass hierarchy. Assuming NH is the truth, we try two fits using NH spectrum and IH spectrum respectively and get two corresponding minimal $\chi^2$, namely $\chi^2_{\mathrm{min}}(\mathrm{NH})$ and $\chi^2_{\mathrm{min}}(\mathrm{IH})$. A discriminator can be defined as
\begin{equation}
\Delta \chi^2=\chi^2_{\rm min}(\rm IH)-\chi^2_{\rm min}(\rm NH),
\end{equation}
A median sensitivity~\cite{Blennow:2013oma}, approximatly $\sqrt{\Delta\chi^2}\sigma$, is often used when comparing the sensitivity of different experiments.

\begin{figure}
\begin{center}
\includegraphics*[width=0.48\textwidth]{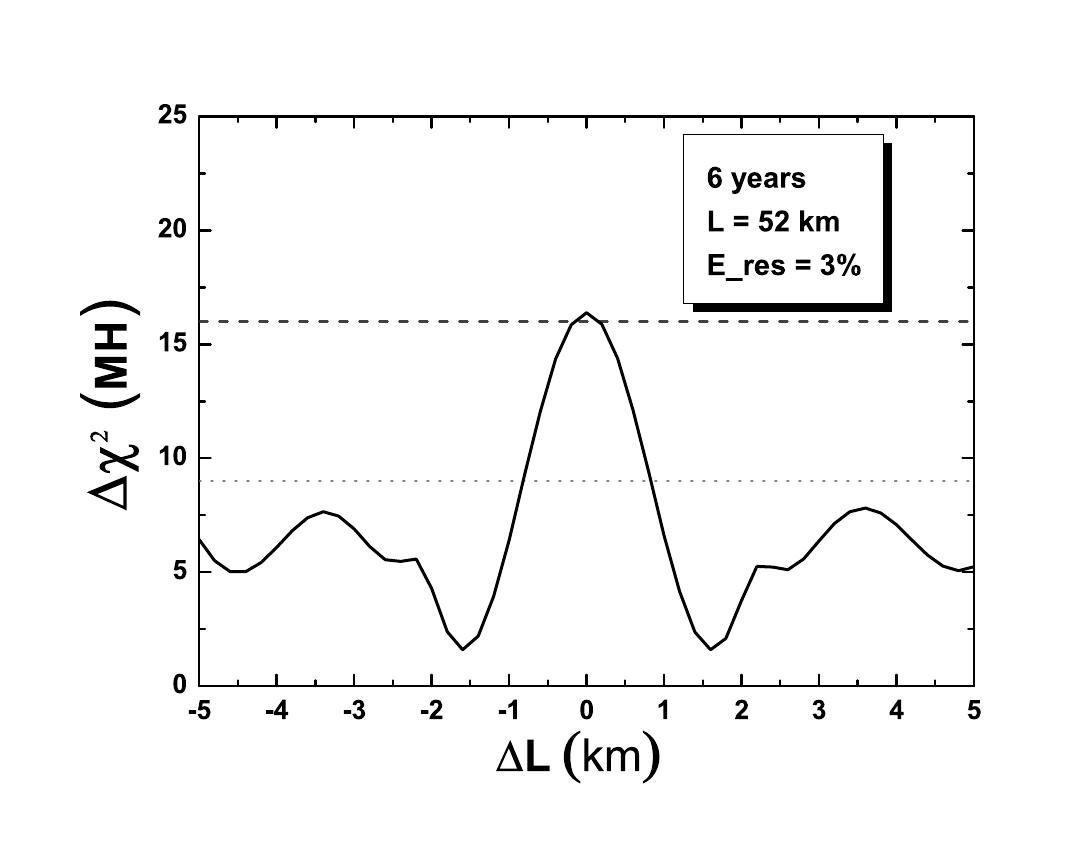}
\end{center}
\caption{The variation (left panel) of the MH sensitivity as a function of the baseline difference of two reactors.\label{fig:disL}}
\end{figure}

\begin{figure}
\begin{center}
\includegraphics*[width=0.48\textwidth]{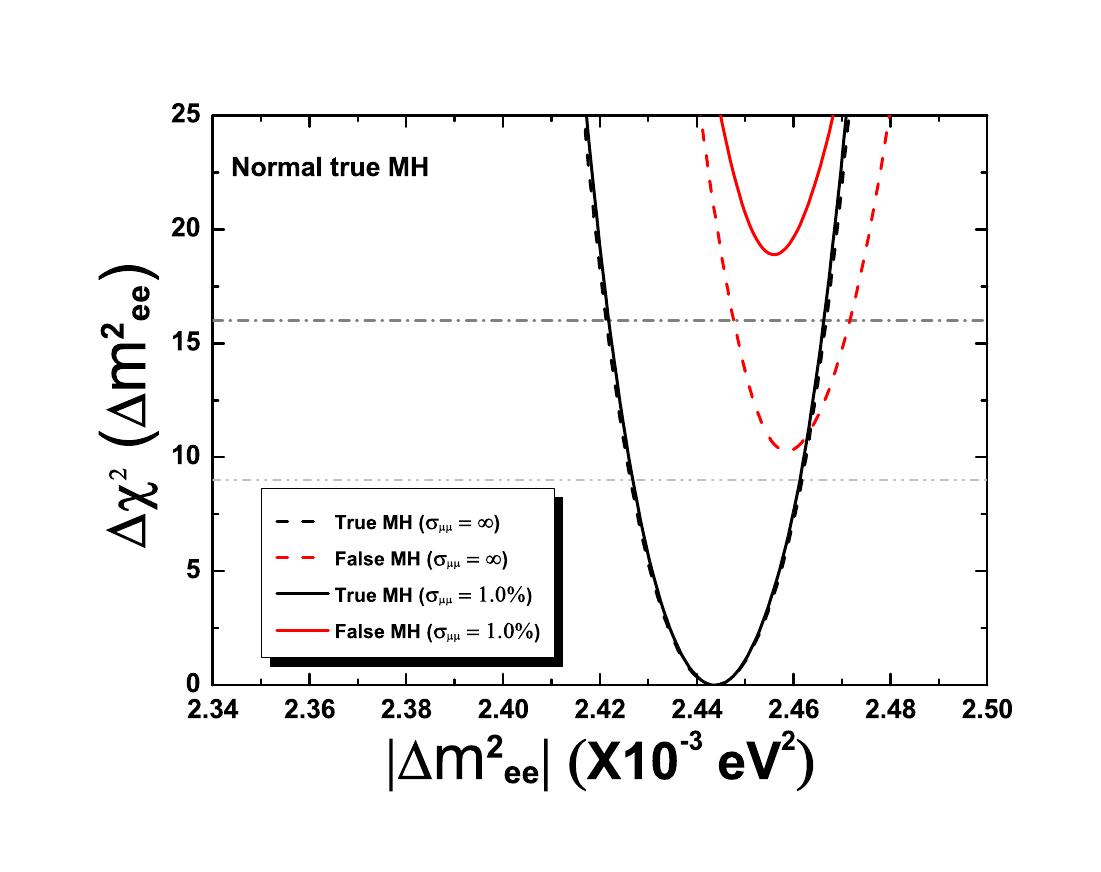}
\end{center}
\caption{Mass hierarchy sensitivity of JUNO. The vertical difference of the curves for the true and false MHs is $\chi^2$ . The solid and dashed lines are for the analyses with and without the prior measurement of $\Delta m^2_{\mu\mu}$. \label{fig:mass-squared}}
\end{figure}

Using the nominal reactor setup list in Table~\ref{tab:reactors} and a 20~kton liquid scintillator detector with a $3\%/\sqrt{E_{vis}}$ energy resolution, we get $\Delta\chi^2 \simeq 11$ ( $\sim3.3\sigma$ at median sensitivity ) by a 6 year running. The importance of the site selection is shown in Figure~\ref{fig:disL}. Assuming two groups of reactors, the sensitivity decreases fast if the baselines are not equal. As a result, $\Delta\chi^2 $ is degraded from 16 at an ideal site with equal baselines to 11 at the candidate site. We find the uncertainties of  $\sin^22\theta_{13}$, $\sin^2\theta_{12}$, $\Delta m^2_{21}$ have slight impacts to the sensitivity. However, the uncertainty of effective mass-squared difference $\Delta m^2_{\mu\mu}$ ( combination of $\Delta m^2_{31}$ and $\Delta m^2_{32}$ ) has large impact to the sensitivity. $\chi^2 \simeq 19$ will be achieved by using a 1\% uncertainty of  $\Delta m^2_{\mu\mu}$ from a prior measurement, shown in Figure~\ref{fig:mass-squared}.

\section{Other physics goals}
Another important goal of JUNO is precision measurement of mixing parameters.  Three mixing parameters, $\Delta m^2_{21}$, $\Delta m^2_{31}$ and $\sin^2\theta_{12}$ can be measured with a precision better than 1\% by precisely measuring the reactor antineutrino oscillation spectrum. It will contribute to the unitarity test of PMNS matrix.

Additionally, JUNO can detect more than 4000 IBD reactions assuming a typical supernova burst at the galaxy center. The energy spectrum of electron antineutrinos ($\bar{\nu_e}$) from supernova can be precisely measured. Comparing with water Cerenkov detector, JUNO has smaller energy threshold (below 1 MeV) and thus can also detected hundreds of neutrino scattering events on protons, $\nu + p \rightarrow \nu + p$. Thus, spectrum information for other neutrino flavors than $\bar{\nu_e}$ can be probed.

Other important goals include observation of geo-neutrinos, solar neutrinos and atmospheric neutrinos, and so on.

\section{Detector design concepts}
\begin{figure}
\begin{center}
\includegraphics*[width=0.45\textwidth]{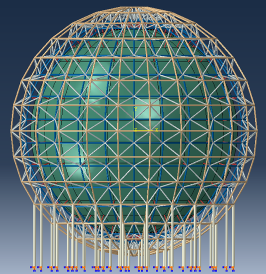}
\end{center}
\caption{A conceptal design for the center detector.\label{fig:detector1}}
\end{figure}

As a preliminary design, JUNO detector includes a central detector to detect reactor antineutrinos and a veto detector to reject cosmic muons and induced backgrounds. Figure~\ref{fig:detector1} is the default design option for center detector. The inner 35~m diameter acrylic vessel contains 20~kton liquid scintillator as target to detect reactor antineutrinos. The outer 40~m diameter stainless steel structure holds $\sim 17000$ 20 inch PMTs to collect the scintillation and Cerenkov lights produced by the interactions of antineutrinos in the liquid scintillator. The acrylic vessel and the stainless steel structure are immerged in a water pool. The water pool is filled with purified water and instrumented with PMTs to function as Cerenkov detector to veto the cosmogenic backgrounds. Above the water pool, a muon tracking detectors probably made by plastic scintillator functions as another veto detector to reject cosmogenic backgrounds.

A backup option is also under R\&D. The inner vessel is a balloon filled by liquid scintillator and supported by acrylic plates.  An outer stainless steel vessel contains mineral oil as shielding and buffer. PMTs are mounted inside of the stainless steel vessel, which is immerged in the water pool. The energy resolution, radioactivity level and the technical challenges are main concerns for the detector R\&D.

\section{Technical challenges}
One key requirement for JUNO detector is the $3\%/\sqrt{E_{vis}}$ energy resolution which is dominated by the photoelectron (PE) yield, namely the number of PEs collected by the PMTs per 1~MeV visible energy ($E_{vis}$) produced by the antineutrino interactions in liquid scintillator. To achieve 3\%/$\sqrt{E_{vis}}$ energy resolution, an unprecedented PE yield of 1100~PEs/MeV is required. The PE yield can be modeled as $\epsilon \cdot C\cdot Y\cdot e^{-d/L}$, where $\epsilon$ is the quantum efficiency (QE) and collection efficiency of converting one optical photon to one PE for PMT, $C$ is the PMT coverage, $Y$ is the scintillation light yield, $L$ is the light attenuation length of the scintillator, $d$ is the transportation distance of the scintillation light. For a large 20~kton detector, better performance of liquid scintillator and PMTs are important to achieve better energy resolution.

For the default design as shown in Figure~\ref{fig:detector1}, the PMT coverage can reach to 75-80\% by careful arrangement of PMT spacing by minimizing the clearance. R\&D efforts are in progress to achieve better performance of liquid scintillator and PMT. The attenuation length of liquid scintillator can be improved from 15~m to 20~m by improving raw material, production process and purification by water extraction, filtration and distillation. A new type of PMT, namely MCP-PMT, can enhance the efficiency by 50-100\% compared with the traditional PMT.  The high QE PMT using SBA photocathode (such as Hammamatzu R5912-100 ) with a 35\% QE is an alternative option.

\begin{figure}
\begin{center}
\includegraphics*[width=0.45\textwidth]{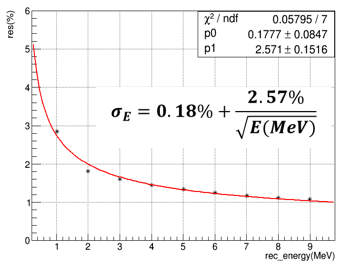}
\end{center}
\caption{Energy resolution by simulation data using a total charge-based energy reconstruction with an ideal vertex reconstruction.\label{fig:Eres}}
\end{figure}

The $3\%/\sqrt{E_{vis}}$ energy resolution is possible to achieve, as demonstrated by a Geant4 detector simulation. Figure~\ref{fig:Eres} shows the result using a setup of 77\% coverage, 35\% QE, and 20~m attenuation length in the simulation software, which has been tuned based on Daya Bay experiment data.

\section{Summary}
JUNO experiment is aimed at neutrino mass hierarchy determination. Using a 20~kton liquid detector with a $3\%/\sqrt{E_{vis}}$ energy resolution, the sensitivity is $3\sigma-4\sigma$ by 6 year running. Additionally, three mixing parameters can be measured with a precision better than 1\%. Other rich physics goals include observation of supernova neutrinos, geo-neutrinos, solar neutrinos, and atmospheric neutrinos. The physics potential is strong, meanwhile the technical challenges are significant. The JUNO project was approved by Chinese Academy of Sciences in 2013 and is planned to be in operation in 2020.




\nocite{*}
\bibliographystyle{elsarticle-num}
\bibliography{martin}



\end{document}